\documentclass[traditabstract, rnote]{aa}
\usepackage{natbib}
\bibpunct{(}{)}{;}{a}{}{ , }
\usepackage[dvipdfm]{graphicx}
\usepackage{amsmath}
\usepackage{booktabs}
\usepackage{txfonts}

\begin{document}
\title{On the possibility for constraining cosmic topology from the celestial distribution of astronomical objects}
\date{Received 4 April 2011 / Accepted 12 March 2011}
\author{H. Fujii \& Y. Yoshii}
\institute{Institute of Astronomy, School of Science, University of Tokyo, 2-21-1, Osawa, Mitaka, Tokyo 181-0015, Japan}

\abstract{
We present a method to constrain cosmic topology from the distribution of astronomical objects projected on the celestial sphere. This is an extension of the 3D method introduced in Fujii \& Yoshii (2011) that is to search for a pair of pairs of observed objects (quadruplet) linked by a holonomy, i.e.,  the method we present here is to search for a pair of celestial sphere $n$-tuplets for $n \geq 3$. We find, however, that this method is impractical to apply in realistic situations due to the small signal to noise ratio. We conclude therefore that it is unrealistic to constrain the topology of the Universe from the celestial distribution, and the 3D catalogs are necessary for the purpose.}

\authorrunning{H. Fujii \& Y. Yoshii}
\titlerunning{The possibility for constraining cosmic topology from the 2D distribution of objects}
\maketitle

\section{Introduction}

Determining the shape of space is one of the most important themes in modern cosmology. Together with the assumption of the cosmological principle, Einstein's General Relativity connects a curvature of space with the average energy density of the universe, and recent observations suggest a flat $\Lambda$-CDM universe (e.g., $\Omega_{\mathrm{tot}}=1.0050^{+0.0060}_{-0.0061}$ from \begin{it}WMAP\end{it}+BAO+SN data, by Hinshaw et al. 2009).

However, while General Relativity describes local geometry (curvature), it predicts nothing about global property of space, namely, topology of the universe. Global topology of the Universe is often assumed to be simply connected with no evidence, though it can be rather multiconnected. 
A multiconnected space with a nontrivial topology can be imagined as a $2K$-polyhedron, called Dirichlet domain, whose $K$ pairs of faces are glued mathematically by holonomies. An object passing through one face immediately returns through the glued face. As a result, multiple images of single objects, often referred to as ``ghosts", appear like those in a kaleidoscope (for detail, see, e.g. Lachi\`eze-Rey and Luminet 1995).

Many methods for constraining cosmic topology are based on this prediction, e.g., \begin{em}circles-in-the-sky\end{em} method (Cornish et al. 1998) is to search for intersections of the last-scattering surface  and the faces of our Dirichlet domain. They are circles with the same temperature fluctuation pattern in the CMB map, because they are copies of a physically identical region. Various authors have searched for matched circles using the \begin{em}WMAP\end{em} data, and obtained diverse results. For example, Aurich (2008) has found a hint of a 3-torus topology, while Roukema et al. (2008) have found a signature of a Poincar\'e dodecahedral space topology. Contrary to them, Cornish et al. (2004), Key et al. (2007), and Bielewicz \& Banday (2011) have found no topological signatures and obtained the lower limit of the size of our Universe. The most recent constraint is $\sim 27.9$ Gpc by Bielewicz \& Banday (2011). These disagreements suggest an existence of methodological problems and bring our interests to the 3D methods, i.e., those that are to use distributions of galaxies, galaxy clusters, or active galactic nuclei.

3D methods are to search for unusual positional patterns of objects in a given astronomical catalog, due to an existence of topological copies (e.g., Lehoucq et al. 1996; Roukema 1996; Uzan et al. 1999; Marecki et al. 2005; Fujii \& Yoshii 2011). For example, if we live in a 3-torus universe, a pair separation histogram (PSH) of a given catalog will show sharp spikes at the separations corresponding to its fundamental lengths (Lehoucq et al. 1996). Fujii \& Yoshii (2011) have introduced a new method that is much more sensitive to topological signatures than the prior ones. This method can apply to any of 17 multiconnected Euclidean spaces, even if the whole universe is comparable to the observed region in size, i.e., in a case that there are only a few topological ghosts.

Our 3D method and others all require spectroscopical observations to calculate the radial distances of objects. A wide field spectroscopic survey is usually difficult, especially for high redshift objects in which we are most interested. Mainly in order to explore CMB data, Bernui \& Villela (2006) have extended the PSH method to the pair angular separation histogram (PASH) method that does not require spectroscopic data when applied to astronomical objects. Their method, however, is so insensitive to topological signatures that the signal does not appear unless we average a number of different but statistically equivalent PASHs, which is almost impossible in practice. 

A significant number of high redshift objects (galaxies or quasars) are recently found, though most of them are not spectroscopically observed yet. This is a good point in time to examine whether these objects are usable for constraining cosmic topology or not. In section 2, we introduce another 2D method to constrain cosmic topology from the distribution of objects projected on the celestial sphere, which is an extension of the 3D method of Fujii \& Yoshii (2011). In section 3, we show that the method is in principle applicable, but in a realistic situation it is no longer useful, similarly to that of Bernui \& Villela (2006). We conclude therefore that spectroscopically observed 3D catalogs are necessary, in order to reveal the topology of the Universe from the distribution of astronomical objects. Throughout the paper we consider flat universes with zero curvature.

\section{Method}

In this section we describe the methodology for judging whether a given pair of $n$-tuplets ($2n$-tuplet) are linked by some holonomy or not. Our assumption is that the universe has zero curvature (Euclidean geometry), and the distances from us to the $2n$ objects are unknown while their celestial positions are known.

A convenient way of writing the holonomies is to use a 4D coordinate system $(w,x,y,z)$ where the simply connected 3-Euclidean space is represented as a hyperplane $w=1$ (see, e.g., Fujii \& Yoshii 2011). Then every holonomy $\gamma$ in a flat universe can be written as a 4D matrix, $\gamma = \gamma_{\mathrm{T}} \gamma_{\mathrm{NT}}$, where $\gamma _{\mathrm{T}}$ and $\gamma_{\mathrm{NT}}$ are a translational part and a nontranslational part, respectively. If a pair of $n$-tuplets $(\vec x_1, \cdots, \vec x_n)$ and $(\vec x'_1, \cdots, \vec x_n')$ are linked by $\gamma$, we have
\begin{equation}
\vec x'_1 = \gamma \vec x_1 = \gamma _{\mathrm{T}} \gamma_{\mathrm{NT}} \vec x_1 = \gamma _{\mathrm{NT}} \vec x_1 + \vec L, \nonumber 
\end{equation}
\begin{equation}
\cdots 
\end{equation}
\begin{equation}
\vec x'_n = \gamma \vec x_n = \gamma _{\mathrm{T}} \gamma_{\mathrm{NT}} \vec x_n = \gamma _{\mathrm{NT}} \vec x_n + \vec L, \nonumber
\end{equation}
where $\vec L$ is the translational vector. Note that a vector $\vec X$ is 4D, $\vec X=(1,X,Y,Z)$, but the 3D part $(X,Y,Z)$ is important. We know all the mathematical possibilities for $\gamma_{\mathrm{NT}}$ (an identity, an $n$-th turn rotation for $n=2, 3, 4,$ or $6$, or a reflection), so these $3n$ equations have $2n+3$ unknown quantities: the distances from us to the $2n$ objects and $\vec L$. If the distances are known from spectroscopic observation, then the unknown quantities are just three, $\vec L=(L_1, L_2,L_3)$, so we derive $3n-3$ conditions as we did for $n=2$ in Fujii \& Yoshii (2011). In that work we introduced a new method that is to search for quadruplets satisfying the condition
\begin{equation}
\vec x_1' - \vec x_2 ' = \gamma_{\mathrm{NT}} (\vec x_1 - \vec x_2),
\end{equation}
and showed that the method is extremely sensitive to topological signatures in a given catalog. 

In the case considered here, however, we do not know the distances and have to take another way. For this, we first denote the 3D positions of objects as follows:
\[ \vec x = \left( \begin{tabular}{c}$x$  \\$y$ \\  $z$  \end{tabular} \right )= \left( \begin{tabular}{c}$r \hat x$  \\$r \hat y$ \\  $r \hat z$  \end{tabular} \right ) =r {\vec R},\]
where $r$ is the unknown distance from us to the object, and the vector $\vec R=(\hat x, \hat y, \hat z)$ represents the known celestial position such that $\hat x^2 + \hat y^2 + \hat z^2 =1$. For the simplest case where $n=3$ and $\gamma_{\mathrm{NT}} = id$, i.e., $\gamma$ is a translation, the following relations hold by eliminating $\vec L$:
\begin{equation}
r_1 \vec R_1 - r_2 \vec R_2 = r_1 ' \vec R_1 ' - r_2 ' \vec R_2 ',
\end{equation}\begin{equation}
r_1 \vec R_1 - r_3 \vec R_3 = r_1 ' \vec R_1 ' - r_3 ' \vec R_3 ',
\end{equation}
Then eliminating $r_2, r_3, r_2 ',$ and $r_3'$ from these equations gives
\begin{eqnarray}
&\bigl[(\vec R_1 \times \vec R_2)_z \hat z_2 ' +(\vec R_2 \times \vec R_2 ')_z \hat z_1 +(\vec R_2 ' \times \vec R_1)_z \hat z_2 \bigr]r_1 +&\nonumber \\
 &\bigl[(\vec R_1' \times \vec R_2 ')_z \hat z_2  +(\vec R_2 ' \times \vec R_2 )_z \hat z_1 ' +(\vec R_2  \times \vec R_1 ')_z \hat z_2 ' \bigr]r_1 ' =0,& \\
&\bigl[(\vec R_1 \times \vec R_3)_z \hat z_3 ' +(\vec R_3 \times \vec R_3 ')_z \hat z_1 +(\vec R_3 ' \times \vec R_1)_z \hat z_3 \bigr]r_1 +&\nonumber \\
&\bigl[(\vec R_1' \times \vec R_3 ')_z \hat z_3  +(\vec R_3 ' \times \vec R_3 )_z \hat z_1 ' +(\vec R_3  \times \vec R_1 ')_z \hat z_3 ' \bigr]r_1 ' =0.&\end{eqnarray}
These are simultaneous equations for $r_1, r_1'$ and should have a solution $(r_1, r_1') \neq (0, 0)$, so the following relation must hold:
\begin{equation}
AD-BC=0, \end{equation}where\begin{eqnarray}
A=(\vec R_1 \times \vec R_2)_z \hat z_2 ' +(\vec R_2 \times \vec R_2 ')_z \hat z_1 +(\vec R_2 ' \times \vec R_1)_z \hat z_2 , \\
B=(\vec R_1' \times \vec R_2 ')_z \hat z_2  +(\vec R_2 ' \times \vec R_2 )_z \hat z_1 ' +(\vec R_2  \times \vec R_1 ')_z \hat z_2 ' , \\
C=(\vec R_1 \times \vec R_3)_z \hat z_3 ' +(\vec R_3 \times \vec R_3 ')_z \hat z_1 +(\vec R_3 ' \times \vec R_1)_z \hat z_3 , \\
D=(\vec R_1' \times \vec R_3 ')_z \hat z_3  +(\vec R_3 ' \times \vec R_3 )_z \hat z_1 ' +(\vec R_3  \times \vec R_1 ')_z \hat z_3 ' . \end{eqnarray}
This relation is written by the celestial positions of the objects, and thus spectroscopic observations are not needed to use it. Though we have considered here the special case that $\gamma$ is a translation, $A, B, C, $ and $D$ for other cases are calculated similarly. If $\gamma$ is a half-turn corkscrew motion or a glide reflection, 
\begin{eqnarray}
A&=&-(\vec R_1 \times \vec R_2)_z \hat z_2 ' +(\vec R_2 \times \vec R_2 ')_z \hat z_1 +(\vec R_2 ' \times \vec R_1)_z \hat z_2 ,  \\
B&=&(\vec R_1' \times \vec R_2 ')_z \hat z_2  -(\vec R_2 ' \times \vec R_2 )_z \hat z_1 ' -(\vec R_2  \times \vec R_1 ')_z \hat z_2 ' , \\
C&=&-(\vec R_1 \times \vec R_3)_z \hat z_3 ' +(\vec R_3 \times \vec R_3 ')_z \hat z_1 +(\vec R_3 ' \times \vec R_1)_z \hat z_3 ,  \\
D&=&(\vec R_1' \times \vec R_3 ')_z \hat z_3  -(\vec R_3 ' \times \vec R_3 )_z \hat z_1 ' -(\vec R_3  \times \vec R_1 ')_z \hat z_3 ' . 
\end{eqnarray}
And if $\gamma$ is an $n$-th turn corkscrew motion for $n=4,3,$ or $6$, then
\begin{tiny}
\begin{eqnarray}
A=(\vec R_1 \times \vec R_2)_z \hat z_2 ' +\{ (\vec R_2 \times \vec R_2 ' )_z \hat z_1 + (\vec R_2 ' \times \vec R_1)_z \hat z_2 \}c + \{ \vec R_2 ' \times (\vec R_2 \times \vec R_1) \} _z s, \\
B=(\vec R_1' \times \vec R_2 ')_z \hat z_2 +\{ (\vec R_2' \times \vec R_2 )_z \hat z_1' + (\vec R_2  \times \vec R_1')_z \hat z_2 '\}c- \{ \vec R_2  \times (\vec R_2 ' \times \vec R_1 ') \} _z s, \\
C=(\vec R_1 \times \vec R_3)_z \hat z_3 ' +\{ (\vec R_3 \times \vec R_3 ' )_z \hat z_1 + (\vec R_3 ' \times \vec R_1)_z \hat z_3 \}c+ \{ \vec R_3 ' \times (\vec R_3 \times \vec R_1) \} _z s, \\
D=(\vec R_1' \times \vec R_3 ')_z \hat z_3 +\{ (\vec R_3' \times \vec R_3 )_z \hat z_1' + (\vec R_3  \times \vec R_1')_z \hat z_3 '\}c- \{ \vec R_3  \times (\vec R_3 ' \times \vec R_1 ') \} _z s,
\end{eqnarray}
\end{tiny}
where $c= \cos(2\pi/ n)$ and $s=\sin( 2 \pi /n)$. Note that the cases for $n=1$ and $n=2$ corresponds to that of translation and half-turn corkscrew motion, respectively.

Given a catalog with $N$ objects, we count the number of sextuplets ($2n$-tuplets for $n=3$) satisfying the condition of $AD-BC=0$, for each type of holonomies, but within a chosen tolerance $\varepsilon$, i.e., $|AD-BC|<\varepsilon$. The number of such sextuplets will be larger for a multiconnected space than stochastically expected for a simply connected one. We assign each object $\vec x_i$ an integer $s_i$, the number of sextuplets that satisfy the condition $|AD-BC|<\varepsilon$ and also include $\vec x_i$ as one of their members.  An $s_i$-histogram for a multiconnected space will have some bumps in the large-$s_i$ region, since topological copies contribute to signal more frequently.

Similar calculations can be done for $n\geq4$ where we have more additional conditions, however, the calculation  time is roughly proportional to ${_N C}_n = N!/(N-n)!n!$, and the calculation for a large $n$ is unrealistically time-consuming. In the next section we show the results for $n=3$, the minimum value above which this method can apply, and discuss the possibility that we can constrain the topology of the Universe by the method.

\section{Simulations and discussions}

In order to see the applicability of the method described in the previous section, we generated toy catalogs in simply and multiconnected  Euclidean spaces. As multiconnected spaces, we considered the following six cases:
\begin{itemize}
\item one with a pair of translations,
\item four with a pair of $n$-th turn corkscrew motions for $n=2,3,4,$ and $6$,
\item and one with a pair of glide reflections.
\end{itemize}
For each simulation, the observed region is a unit sphere centered at the observer's position, and the translational distance is $L=|\vec L|=1.4$, which implies that $1 - \int_{-0.7} ^{0.7} \pi (1-x^2)dx / \frac{4\pi}{3} \sim12$\% of the observed region is a copy of some part of the ``first" copy of the universe. Now consider $L$ to be $c/H_0 \simeq 4.2$ Gpc, where $c$ and $H_0$ are the speed of light and the Hubble parameter, respectively. With this example scale, the radius of the observed region is $\sim3.0$ Gpc, corresponding to $z \sim 0.9$. The effects of global inhomogeneity that had been  investigated in detail in Fujii \& Yoshii (2011) were not considered here: the axes for the $n$-th turn corkscrew motions and the reflectional plane for the glide reflections pass through the observer, and we chose the correct coordinate systems.

Our toy catalogs were made as follows: we uniformly distributed 50 real objects, and then their copies were generated for each multiconnected space. The total number of objects $N$ (50 real ones plus their copies) is different from simulation to simulation, since some of the copies were beyond the observed region. Next, another catalog in the simply connected space with the same number of objects was made for comparison. All objects were projected on the celestial sphere, and sextuplets satisfying the condition $|AD-BC|<\varepsilon$ were searched for. 

\begin{figure}[!htb]
\centering
\resizebox{\hsize}{!}{\includegraphics{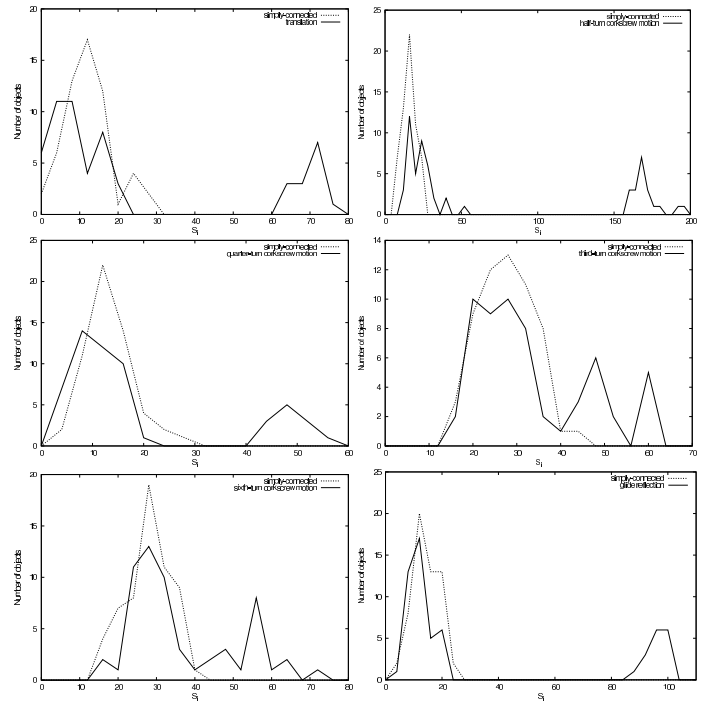}}
\caption{The $s_i$-histograms for various types of holonomies: translation (\emph{top left}), half-turn corkscrew motion (\emph{top right}), quarter-turn corkscrew motion (\emph{mid left}), third-turn corkscrew motion (\emph{mid right}), sixth-turn corkscrew motion (\emph{lower left}), and glide reflection (\emph{lower right}). Bumps constituted by topological copies for these types are seen in the large $s_i$ regions for multiconnected spaces.} 
\label{figure1} 
\end{figure}

First we succeeded in detecting topological signal with the choice of $\varepsilon =10^{-8}$. The results are given in Table \ref{table1}. The number of sextuplets that satisfy the condition of $|AD-BC| < \varepsilon$, for each case, is given as ``topological index". [Note that we had normalized the topological indices with respect to the total number of quadruplets in Fujii \& Yoshii (2011), but not here.] As expected, topological index for a multiconnected space is larger than that for the simply connected one, which successfully distinguishes the two topologies. The difference can more clearly be seen in $s_i$-histograms as shown in Figure \ref{figure1}. Some bumps are seen in the large-$s_i$ regions for multiconnected cases, because of the topological copies that contribute to the signal more frequently than  a stochastic expectation.

These results show that this 2D method in principle constrains the topology of the Universe, however, the practical application of the method is not so straightforward since the condition considered here is optimistic and somewhat arbitrary:
\begin{itemize}

\item The tolerance as small as $\varepsilon = 10^{-8}$ cannot be chosen when correctly considering the limited resolution, e.g.,  with the typical ground-based optical instruments, $\lesssim 1$ arcsec, which implies the uncertainties in $\varepsilon$ of the order of $10^{-6}$. Moreover, even if we have an ideal instrument, due to the peculiar velocity, the quantity $|AD-BC|$ will be larger than $10^{-8}$ even for a really topological sextuplet. For example, the typical peculiar velocity of $v=500$ km/sec and the time lag of $10^8$ yr (corresponding to an upper limit to quasar lifetimes) implies that $\varepsilon$ should be larger than $\sim 10^{-5}$ at the redshift $z \sim 5$. These large tolerances drastically enhance the stochastic noise.
\item The ratio of the number of topological ghosts to the total number of objects ($N$) is large here, $\sim 12-17 \%$. In reality, however, it can be much smaller than this because of a tiny number of ghosts, a large number of objects, or the both, so the topological signal will be hidden by the stochastic noise: roughly speaking, the former scales as $N^3$ while the latter scales as $N^6$. 
\item Each simulation here took several minutes using an ordinary personal computer. Howeover, calculations for a realistic value of $N \sim 10^3$ takes an extremely long time that is roughly proportional to ${_N C}_3 = N(N-1)(N-2)/6$.
\end{itemize}
For example, with the choice of $\varepsilon =10^{-5}$ and $N \sim 60$, the stochastic noise has an order of $10^5$, completely hiding the topological signal of $10^2$. The situation gets even worse for a larger $N$.

Thus the method will be impractical in realistic situations due to small signal to noise ratio. One way to overcome the problem may be using a large $n\geq4$, rather than $n=3$. For a given pair of $n$-tuplets to be linked by a holonomy, there are $n-2$ conditions to be satisfied, thus we have $n-2$ filters. The possibility $P(n)$, such that a non topological $2n$-tuplet passes through all the filters by chance, is monotonically decreasing with $n$. Therefore, the stochastic noise $P(n) \times {_N C}_n$, can be suppressed if we choose a sufficiently large $n$. However, as mentioned in the previous section, the calculation time roughly scales as ${_N C}_n = N!/(N-n)!n!$ and calculations for such a large $n$ are extremely time-consuming.

Considering these circumstances, we must conclude that it is unrealistic to constrain cosmic topology from the celestial 2D distribution of objects. Rather, for the practical purpose, we need the 3D catalogs of cosmic objects obtained by spectroscopic observations. The new crystallographic method introduced by Fujii \& Yoshii (2011) can be applied to such catalogs, and provides us a knowledge about the shape of the Universe. This will more than ever strengthen the motivation for promoting a large-scale spectroscopy survey of high redshift objects.

\begin{table*}[!htb]
\centering
\begin{tabular}{c|c|c|c|c} \hline \hline
Holonomy type & Number of  & Number of  & Topological & Topological \\ 
& quasars & ghosts & index $S_{\mathrm{mult}}$ & index $S_{\mathrm{simp}}$ \\ \toprule
Translation & 57 & 14 & 224 & 116 \\
Half-turn corkscrew motion & 60 & 20 & 733 & 175 \\ 
Quarter-turn corkscrew motion & 58 & 16 & 186 & 138 \\
Third-turn corkscrew motion & 57 & 14 & 318 & 235 \\
Sixth-turn corkscrew motion & 59 & 18 & 372 & 291\\
Glide reflection & 58 & 16 & 359 & 153\\ \bottomrule
\end{tabular}
\caption{Results for various types of holonomies. Topological index $S_{\mathrm{mult}}$ is the number of sextuplets satisfying $|AD-BC|<10^{-8}$ for a multiconnected space, and $S_{\mathrm{simp}}$ is for the simply connected one.}
\label{table1}
\end{table*}

\acknowledgement{We thank T. Minezaki, T. Tsujimoto, T. Yamagata, Y. Sakata, T. Kakehata, and K. Hattori for useful discussions and suggestions.}

\nocite{*}
\bibliography{Untitled}
\end{document}